\documentclass[aps,twocolumn,footinbib,floatfix]{revtex4}

\usepackage{amssymb}
\usepackage[pdftex]{graphicx}	
\usepackage[applemac]{inputenc}
\usepackage{amsmath,amssymb,amsthm}
\usepackage{mathbbol,bbm}
\usepackage{graphicx,float}
\usepackage{bm,dsfont}
\usepackage{color}

\def\Tr{ {\rm{Tr }}}

\usepackage{booktabs}

\begin{document}
\title{Topological Heat Transport and Symmetry-Protected Boson Currents}
\author{\'Angel Rivas and Miguel A. Martin-Delgado}
\affiliation{Departamento de F\'{\i}sica Te\'orica I, Universidad Complutense, 28040 Madrid, Spain.\\
CCS -Center for Computational Simulation, Campus de Montegancedo UPM, 28660 Boadilla del Monte, Madrid, Spain.}

\vspace{-3.5cm}

\begin{abstract}
The study of non-equilibrium properties in topological systems is of practical and fundamental importance. Here, we analyze the stationary properties of a two-dimensional bosonic Hofstadter lattice coupled to two thermal baths in the quantum open-system formalism. Novel phenomena appear like chiral edge heat currents that are the out-of-equilibrium counterparts of the zero-temperature edge currents. They support a new concept of dissipative symmetry-protection, where a set of discrete symmetries protects topological heat currents, differing from the symmetry-protection devised in closed systems and zero-temperature. Remarkably, one of these currents flows opposite to the decreasing external temperature gradient. As the starting point, we consider the case of a single external reservoir already showing prominent results like thermal erasure effects and topological thermal currents. Our results are experimentally accessible with platforms like photonics systems and optical lattices.
\end{abstract}

\maketitle

\section*{INTRODUCTION}

Topological insulators represent a new state of matter that have attracted much attention due to their exotic physical properties and its potential applications in spintronics, photonics, etc., that may revolutionize these fields \cite{rmp1,rmp2,Moore,LibroBernevig}.

One of the most active areas in topological insulators is their quantum
simulation with the goal of realizing novel physical properties that
are otherwise very difficult to realize in a standard condensed matter system \cite{JZ,non-abelian,review_OLs_TI}.
Among these quantum simulators, bosonic systems such as ultracold atoms in optical lattices or photonic chips, stand up as versatile and promising experimental platforms that have achieved enormous progress and points towards near-future technological applications \cite{Haldane,Hafezi1,Segev,Hafezi2,MacDonald,Bloch,Ketterle,Spielman}.

Whereas these systems have been extensively studied for the idealized and isolated case, very little is known about the response of these setups to the action of external thermal fluctuations or external dissipation \cite{Instability1D,DMTI,Mazza,Baranov,Linzner,Albert}. This is both true when the system is perturbed by some external heat bath or in an out-of-equilibrium situation where the system interacts with two different heat sources. Then, a natural question that arises is to what extent the topological properties of these systems affect heat currents and transport under these circumstances. In particular, and very importantly, are there new topological heat currents showing exotic behavior?

\begin{figure}[t]
	\includegraphics[width=\columnwidth]{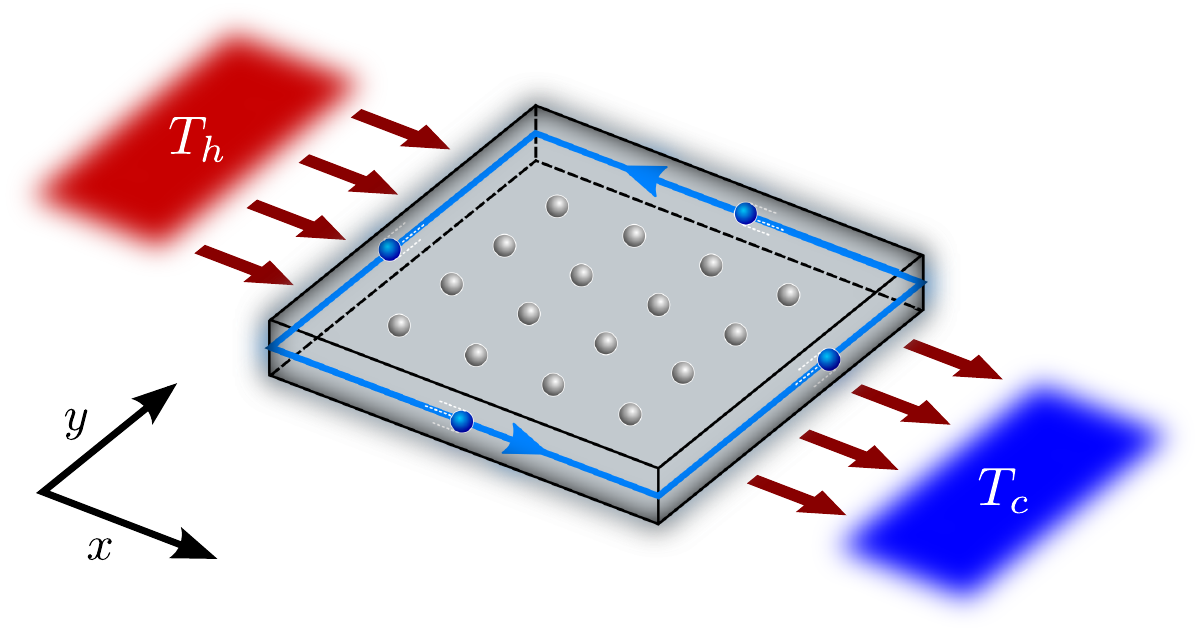}
	\caption{Schematic arrangement considered throughout this work. A Hofstadter boson system is set in contact with two oppositely sited thermal baths at respective temperatures $T_h$ (hot bath) and $T_c$ (cold bath). As a result, exotic chiral currents are induced on the system. The brown arrows show the direction of heat.}
	\label{fig:fig1}
\end{figure}

In this work we have addressed these novel and relevant issues showing that the exposition to thermal sources has not necessarily a detrimental effect. It can actually produce new topological features outside the paradigm of closed systems, that may have also technological applications. Among these results we may highlight:

(i) Thermal Erasure Effect (TEE):
 There exists a wide range of temperatures such that the current becomes
more localized on the edge than for an individually excited edge
mode. In particular, bulk currents are negligible in comparison to the edge ones as
an effect of thermal fluctuations.

(ii) Topological Thermal Currents (TTC): The edge currents
driven in the system by the presence of a single thermal bath
are topologically protected against disorder.

(iii) Edge Nonequilibrium Crosscurrent (ENC): When the system is
in contact with two baths at different temperatures (i.e.
nonthermal equilibrium situation), a chiral current is also
induced so that on one edge of the system the current
flows in opposite direction to the heat. This can be thought of as a local entropy decrease on that edge \cite{Lieb}.
Thus, the violation of the second law (Clausius form) by the nonequilibrium crosscurrent
is just apparent since the net heat flow is from the hottest bath to the coldest one, as it should be.

(iv) Symmetry-Protected Non-Equilibirum Currents (SPNC). As a
difference with standard topological currents, the nonthermal
equilibrium currents present selective robustness.
The edge current is immune to the presence of disorder provided that it satisfies a particular global spatial symmetry.

This notion of SPNC is motivated by a similar notion for topological insulators environmentally isolated (closed systems), but there is one important difference. Namely, the currents here are effectively protected by discrete symmetries as clearly shown in our numerical simulations for a wide range of temperatures. However, at very high temperatures the protection ceases to be operative, as it may be expected. This has mathematical implications, e.g. it is not possible to ascribe a usual Chern number (integer) to these currents as in the standard closed case. 
Nevertheless, this comprises the main novel feature of our study, the concept of \emph{dissipative symmetry-protection}: an effective notion of symmetry-protected heat currents valid for open systems. Moreover, the SPNC are switchable without altering the thermal baths: by controlling the orientation of the external magnetic field its chirality can be modified at will (see Fig.~S2 in the supplementary information document). This may represent a new way to technologically exploit heat flows.

\section*{RESULTS}

\subsection*{System and Dynamics}
We consider a $N\times N$ square lattice of bosonic modes in contact with two sets of local (bosonic) thermal reservoirs, one, on the left, at temperature $T_h$ and other on the right, at temperature $T_c$ (without lost of generality we shall assume $T_h\geq T_c$) as depicted in Fig.~\ref{fig:fig1}. The lattice Hamiltonian is assumed to be
\begin{equation}\label{SysH}
H_S=\sum_{x,y}\hbar \omega_0 a_{x,y}^\dagger a_{x,y}+V,
\end{equation}
with
\begin{align}\label{H1}
V=-\hbar J\sum_{x,y} a_{x+1,y}^\dagger a_{x,y} e^{-2\pi \alpha i y} +a_{x,y+1}^\dagger a_{x,y}+{\rm h.c.}
\end{align}
This is the bosonic version of the Hofstadter model for the integer quantum Hall effect \cite{Butterfly}. This bosonic Hofstadter Hamiltonian has been obtained in controlled systems like ultracold gases in optical lattices subject to laser induced tunneling \cite{Bloch,Ketterle,Spielman}, and photonic circuits arranging differential optical paths  \cite{Hafezi2}. Furthermore, similar dynamics can also be found in photonic crystals \cite{Segev,MIT2}. These systems reproduce the effect of the magnetic flux $\alpha$ by several artificial techniques allowing for the exploration of quantum Hall physics with neutral atoms or photons. The interaction with local reservoirs is modeled by a sum of individual Hamiltonians accounting for a standard quadratic interaction:
\begin{align}\label{HSR1}
H_{SR}=\sum_{j,y} &g_j (A_{j,y}+A^\dagger_{j,y})(a_{1,y}+a^\dagger_{1,y}) \nonumber\\
+ &g_j (B_{j,y}+B^\dagger_{j,y})(a_{N,y}+a^\dagger_{N,y}).
\end{align}
Here, $A_{j,y}$ and $B_{j,y}$ denote bosonic operators of the mode with frequency $\omega_j$ of the reservoir at position $y$, at the left and right hand side (i.e. hot and cold), respectively; and $g_j$ is the coupling constant assumed to be the same for all reservoirs. Although the assumption of local reservoirs is quite natural in the context of individual addressing, where access to each individual site of the lattice is possible, in the practice it turns out to be reasonably valid for a wide variety of physical situations.

Assuming weak system-bath couplings and following the usual steps for the derivation of the master equation (see Methods), we obtain
\begin{widetext}
\begin{align}\label{ME1}
\frac{d\rho}{dt}=-\frac{i}{\hbar}[H_S,\rho] &+ \sum_k \gamma \big\{s_k  [\bar{n}_k(T_h)+1]+r_k [\bar{n}_k(T_c)+1]\big\} \Big(b_k \rho b_k^\dagger - \frac{1}{2}\{b_k^\dagger b_k,\rho\}\Big) \nonumber \\
&+ \sum_k \gamma \big[s_k \bar{n}_k(T_h)+r_k \bar{n}_k(T_c)\big] \Big(b_k^\dagger \rho b_k - \frac{1}{2}\{b_k b_k^\dagger,\rho\}\Big),
\end{align}
\end{widetext}
where $\mathcal{L}$ represents the Liouvillian operator, $b_k$ stands for the normal modes of $H_S$, $\bar{n}_k(T)=\{\exp[\hbar \omega_k/(k_B T)]-1\}^{-1}$ denotes the mean number of bosons with frequency $\omega_k$ and temperature $T$, and $\gamma$ is a constant that depends on the strength of the coupling $\gamma\sim g_j^2$. Furthermore the constants $s_k$ and $r_k$ are related to the coordinates in real space of ``one-particle'' eigenfunctions $\psi_k(x,y)$ of $H_S$, via
\begin{equation}\label{srdef1}
s_k=\sum_{y=1}^N |\psi_k(1,y)|^2 \quad\text{and}\quad r_k=\sum_{y=1}^N |\psi_k(N,y)|^2.
\end{equation}
They correspond to left ($x=1$) and right ($x=N$) sides near the reservoirs, respectively.

We aim at studying the heat current in the asymptotic limit, once the system has reached stability. For the equilibrium situation $T_h=T_c=T$, the decay rate becomes $\gamma(s_k+r_k)$ and the master equation \eqref{ME1} drives the system towards thermal equilibrium with the baths, so that the steady state is $\rho_{\rm ss}:=\lim_{t\rightarrow\infty}\rho(t)=\rho_\beta=\frac{e^{-\beta H_S}}{Z}$ with $\beta=1/(k_B T)$ and $Z=\Tr[\exp(-\beta H_S)]$.
In the general nonthermal equilibrium case $T_h>T_c$, the steady state can be written as $\rho_{\rm ss}=W^{-1} \exp\left(-\sum_{k} \frac{\hbar \omega_k}{k_B T^{\rm eff}_k} b_k^\dagger b_k\right)$ where $W$ is a normalization constant and the quantity $T^{\rm eff}_k$ plays the role of a mode-dependent effective temperature with the form
\begin{widetext}
\begin{equation}\label{Teff}
  T^{\rm eff}_k:=\frac{\hbar \omega_k}{k_B\log\left\{\frac{\exp\big(\tfrac{\hbar \omega_k}{k_B T_h}\big)\big[\exp\big(\tfrac{\hbar \omega_k}{k_B T_c}\big)-1\big] s_k + \exp\big(\tfrac{\hbar \omega_k}{k_B T_c}\big)\big[\exp\big(\tfrac{\hbar \omega_k}{k_B T_h}\big)-1\big] r_k}{\big[\exp\big(\tfrac{\hbar \omega_k}{k_B T_c}\big)-1\big] s_k + \big[\exp\big(\tfrac{\hbar \omega_k}{k_B T_h}\big)-1\big] r_k}\right\}}.
\end{equation}
\end{widetext}

Since the lattice holds a $\Theta\Sigma_y$ symmetry [$\Theta$: time-reversal (changing $\alpha\leftrightarrow-\alpha$), $\Sigma_y$: 2D reflection across the $y$ axis], we have $|\psi_k(1,y)|^2=|\psi_k(N,y)|^2$ and therefore
\begin{equation}\label{skrk}
s_k=r_k.
\end{equation}
As a consequence, in general terms, the physical properties described by the master equation \eqref{ME1} are invariant under all symmetries respecting both $H_{S}$ and Eq.~\eqref{skrk}, and thus the Liouvillian operator $\mathcal{L}$. These are $\Theta\Sigma_y$ and $R_\pi$, a $\pi$-rotation along the orthogonal direction to the lattice [$R_\pi\psi_k(1,y)= \psi_k(N,N+1-y)$]. As commented, for $T_h=T_c=T$, the physics in the stationary limit is independent of $s_k$ and $r_k$ because their effect on the master equation \eqref{ME1} is just a renormalization of $\gamma\rightarrow\gamma(s_k+r_k)$, so that the symmetries of $r_k$ and $s_k$ do not play any significant role. This is consequent with the fact that no original spatial symmetry is broken as no temperature gradient is applied. However, things are different if $T_h>T_c$. Specifically, we can await for robustness of the chiral currents in the Hofstadter model also in the nonthermal equilibrium situation, as least if the spatial distribution of defects remains invariant under $\Theta\Sigma_y$ or $R_\pi$ such that Eq.~\eqref{skrk} is satisfied. If the latter is not the case, the physical properties of the master equation \eqref{ME1} may be affected and there is no guarantee that the robustness of the currents was preserved. Hence, a distribution of defects that changes by $\Theta\Sigma_y$ and $R_\pi$, might destabilize the chiral current. Mathematically, this corresponds to a pair of $\mathbb{Z}_2$ symmetries, 
$\mathbb{Z}_2\equiv\{\mathds{1},R_\pi\}$ and $\mathbb{Z}_2^\ast\equiv\{\mathds{1},\Theta\Sigma_y\}$. Note that different symmetry-protected topological behavior may be expected from the fact that the system cannot be deformed from the thermal to nonthermal situations in a continuous way avoiding symmetry-breaking.

\subsection*{Internal and External Currents}

We shall distinguish between two types of currents, the external ones, which describe the exchange of energy between system and baths, and the internal ones, that concern to the transport inside the $N\times N$ lattice system array. The external currents operators are derived from the master equation \eqref{ME1} and the continuity equation for the total energy (see Methods), and take the form
\begin{align}
  \mathcal{J}_h:=-\hbar \sum_k \omega_k \gamma_k s_k [b_k^\dagger b_k- \bar{n}_k(T_h)],\\
  \mathcal{J}_c:=-\hbar \sum_k \omega_k \gamma_k r_k [b_k^\dagger b_k- \bar{n}_k(T_c)],
\end{align}
for the exchange with hot and cold bath, respectively. In the steady state limit, their expectation values become
\begin{equation}
  \langle\mathcal{J}_{h,c}\rangle_{\rm ss}=\hbar \sum_k \omega_k \gamma_k r_k s_k \left[\frac{\bar{n}_k(T_{h,c})-\bar{n}_k(T_{c,h})}{s_k + r_k}\right].
\end{equation}
Since $T_h\geq T_c$, $\langle\mathcal{J}_h\rangle_{\rm ss}=-\langle\mathcal{J}_c\rangle_{\rm ss}\geq0$ and the heat current abides with the second law flowing from the hot bath to the system and from system to the cold bath.

External currents are less exposed to topological properties (e.g. their values are quite independent of the value of $\alpha$) than internal currents. This is due to the fact that the system-bath coupling $H_{SR}$
does not enjoy any special topological feature. These are present in the intersystem coupling $V$, Eq.~\eqref{H1}, which is directly related to internal currents.

Gauge invariant operators for internal currents are derived in similar fashion from the continuity equation for site populations $\langle a_{x,y}^\dagger a_{x,y}\rangle$ in terms of the total Hamiltonian $H_S+H_{SR}$, they read
\begin{align}\label{IntCurrent1}
  \mathcal{J}_{(\rightarrow x),y}&:=iJ\big(a_{x,y}^\dagger a_{x-1,y}e^{-2\pi \alpha i y}-a_{x-1,y}^\dagger a_{x,y} e^{2\pi \alpha i y}\big),\\
  \mathcal{J}_{x,(\rightarrow y)}&:=iJ\big(a_{x,y}^\dagger a_{x,y-1}-a_{x,y-1}^\dagger a_{x,y}\big),\label{IntCurrent2}
\end{align}
where the subindex $(\rightarrow x)$ is a short notation for $(x-1 \rightarrow x)$. Thus $\mathcal{J}_{(\rightarrow x),y}$ denotes the operator for the current leaving site $(x-1,y)$ and entering in $(x,y)$. Similarly for $(\rightarrow y)$. Note that these internal currents describe just a flow of bosons (instead of energy). At the moment, we neither intend or need to associate any specific energy to the current from some site to its adjacent independently of what normal mode is excited on the lattice.

\subsection*{Thermal Equilibrium Currents}

\begin{figure}[t]
	\centering
	\includegraphics[width=\columnwidth]{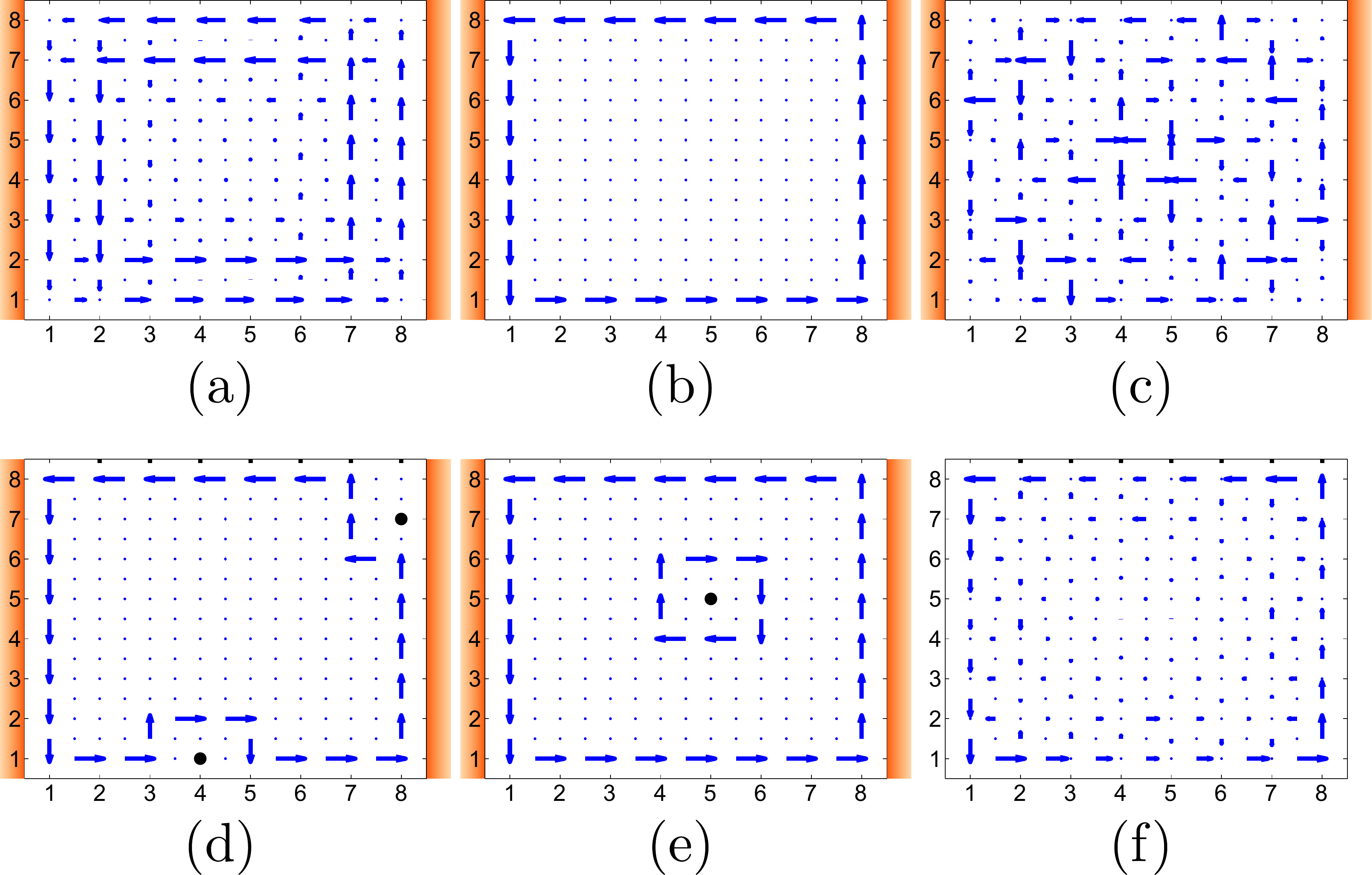}
	\caption{Current patterns for the system in contact with thermal baths at the same temperature $T_h=T_c=T$: (a) for $T=4\ {\rm K}$,  (b) for $T=2500\ {\rm K}$, and (c) for $T=10^{15}\ {\rm K}$; and under the presence of different defects represented with a black dot in (d) and (e) for $T=2500\ {\rm K}$. In (f) the pattern for an individual excited edge mode is depicted. For the sake of illustration the magnitude of the currents has been normalized in every subplot, see main text for specific numbers.}
	\label{Fig:Thermal}
\end{figure}

We shall move our attention to the behavior of the internal currents, refering to the external ones when appropriate. Motivated by realistic simulations \cite{Hafezi2}, we have taken $N=8$ and $\omega_0=193\ {\rm THz}$,  $J=2.6\ {\rm THz}$ and $\alpha=0.15$; different values may be considered leading to similar conclusions. In Figs.~\ref{Fig:Thermal}(a), \ref{Fig:Thermal}(b) and \ref{Fig:Thermal}(c) the current pattern on the lattice in the steady state limit is depicted as a function of bath temperature. For low temperatures ($\lesssim\! 70$ K for this choice of parameters) the ground state, which is a bulk state, is mainly populated and no edge current is observed. As temperature increases (in a range of 70--$10^{13}$ K approximately), the edge states start populating and current start concentrating on the edge [Figs.~\ref{Fig:Thermal}(a) and \ref{Fig:Thermal}(b)]. For high enough temperature everything gets mixed up [Fig.~\ref{Fig:Thermal}(c)].

Notably, the accumulation of current on the edge relative to the bulk is in fact higher for a thermal state at a suited temperature than for a single edge mode, Fig.~\ref{Fig:Thermal}(f). Specifically, for the aforementioned values in Fig.~\ref{Fig:Thermal}, the ratio between edge/bulk current is about 2 for 4 K, $10^3$ for 2500 K,  1 for $10^{15}$ K, and 10 for the individual edge mode (see Fig. S3  in the supplementary information document for a detailed plot). This TEE, the point (i) above, can be related to a combined action. On the one hand, the currents can be written as a sum $\langle\mathcal{J}\rangle=\sum_k \bar{n}_k(T)\langle\mathcal{J}\rangle_k$ with $\langle\mathcal{J}\rangle_k$ the current for one excitation in the mode $k$. It is well known that the density of states of the energy spectrum is much higher for bulk states than for edge states (see Fig. S1 in the supplementary information document). Therefore the factor $\bar{n}_k(T)=\{\exp[\hbar \omega_k/(k_B T)]-1\}^{-1}$ as a function $k$ is flatter for values of $k$ belonging to bulk modes (bulk currents) than for edge modes (edge currents), providing more mixing in the bulk than on the edge. On the other hand, the contribution of the edge states is amplified on the edges since they decay very rapidly into the bulk.

In addition, as anticipated in point (ii) above, we observe that the edge thermal current is robust under the presence of defects (TTC), Figs.~\ref{Fig:Thermal}(d) and \ref{Fig:Thermal}(e). Defects are effectively created by far off-detuning of site local energies. Moreover, if a defect is allocated in the bulk, the system generates a current around it with opposite direction to the edge current Fig.~\ref{Fig:Thermal}(e). In this regard, it is worth to mention this is so despite the bulk has not direct contact with the baths, only left and right edges are in contact with them. Note that the presence of local currents in the equilibrium case does not lead to any thermodynamic inconsistency as the net heat flow as accounted for the external currents vanishes $\langle\mathcal{J}_h\rangle=\langle\mathcal{J}_c\rangle=0$.

\subsection*{Nonthermal Equilibrium Currents}

For the out-of-thermal-equilibrium situation, $T_h>T_c$, an edge chiral current is also found. Actually, we obtain a similar pattern as for baths at the same temperature. However in this case, the current flows on one edge in the opposite direction to the heat. This apparent violation of the second law on one edge is a topological effect (ENC), and, as stated in point (iii) above, it is not a contradiction with thermodynamics as the total heat current as measured by external currents ($\langle \mathcal{J}_h\rangle_{\rm ss}=-\langle \mathcal{J}_c \rangle_{\rm ss}=29.85$ MeV/s for this choice of parameters) does satisfy the second law. Yet, this feature of the edge current is very remarkable as it implies that a local measurement on an edge does not provide enough information to infer the positions of hot and cold baths in this topological system.

\begin{figure}[t]	
	\centering
	\includegraphics[width=\columnwidth]{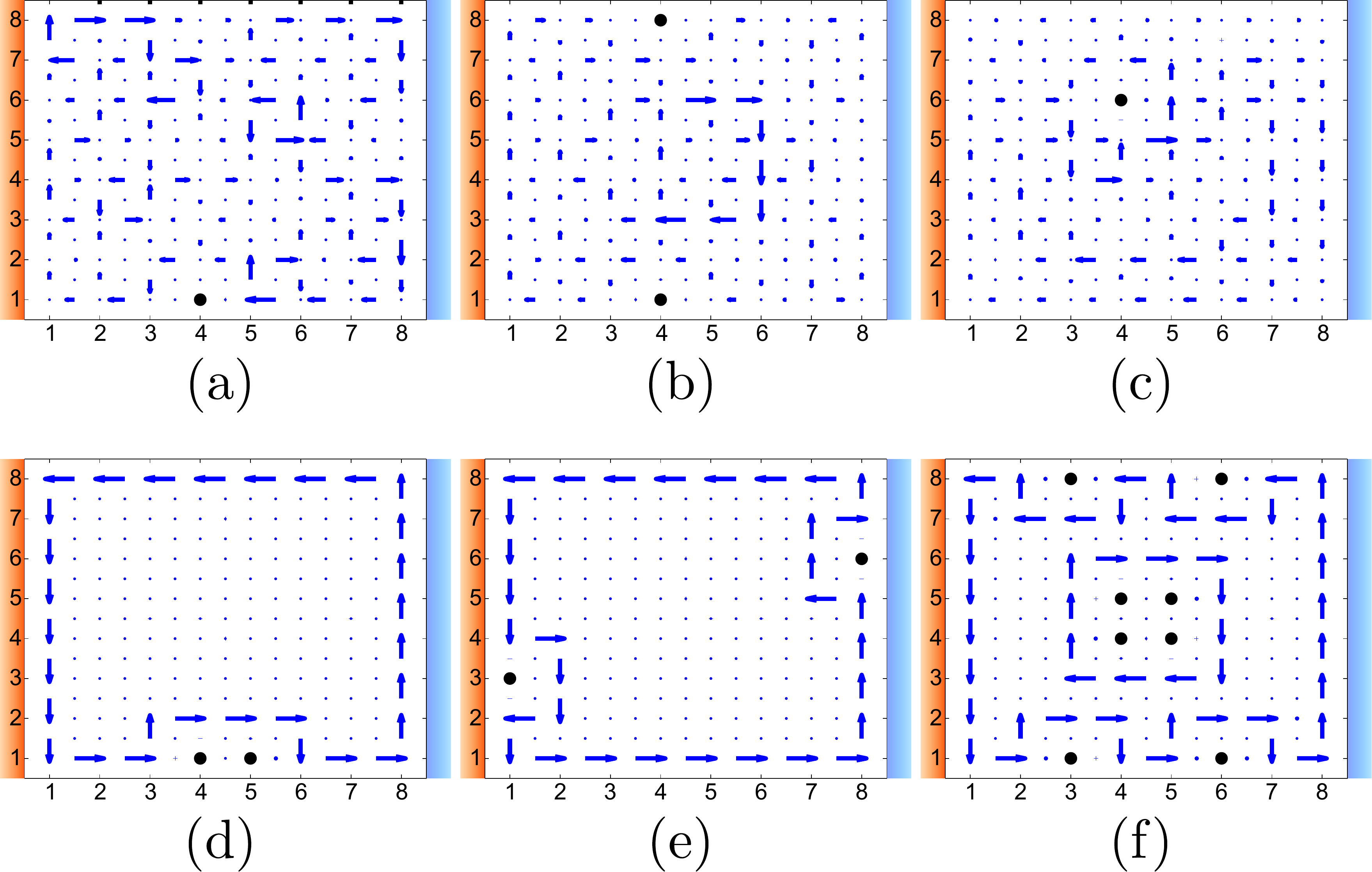}
	\caption{Current patterns for an out-of-equilibrium situation under the presence of defects indicated with a black dot ($T_h=2500$ K and $T_c=1500$ K, similar results are obtained for other values avoiding extremely high and low temperatures, see Fig. \ref{Fig:Thermal}). In (a), (b) and (c) the distribution of defects respects neither $\Theta \Sigma_y$ nor $R_\pi$ symmetries, and the currents become unstable. In contrast, in (d), (e) and (f) the distribution of defects is invariant under $\Theta \Sigma_y$, $R_\pi$, and both $\Theta \Sigma_y$ and $R_\pi$, respectively, and the currents remain robust.}
	\label{Fig:NONThermal2}
\end{figure}

The behavior of the nonthermal equilibrium current under the presence of impurities in the lattice exhibits novel and intriguing features. As aforementioned, just from the symmetries of the master equation \eqref{ME1}, we may expect robustness for a distribution of defects satisfying either $\Theta\Sigma_y$ or $R_\pi$. For a different situation, the result is far from clear. It turns out that the nonthermal equilibrium currents are indeed stable for defect distributions preserving these symmetries, see Figs.~\ref{Fig:NONThermal2}(d), \ref{Fig:NONThermal2}(e) and \ref{Fig:NONThermal2}(f). Nonetheless, currents become unstable for spatially distributed defects not complying with them, Figs.~\ref{Fig:NONThermal2}(a), \ref{Fig:NONThermal2}(b) and \ref{Fig:NONThermal2}(c).

\begin{figure}[t]
	\centering
	\includegraphics[width=\columnwidth]{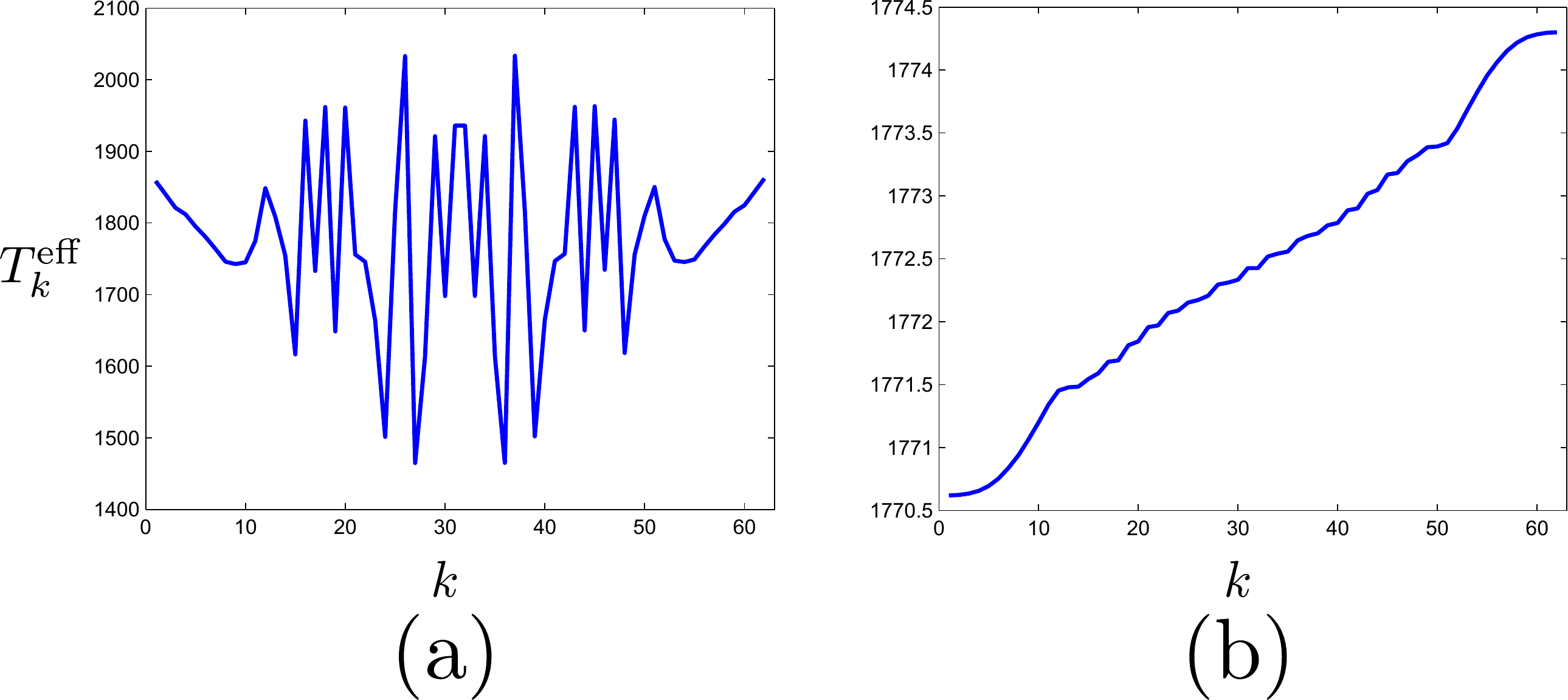}
	\caption{Dependence of $T^{\rm eff}_k$ on the specific mode $k$ for: (a) an arbitrary distribution of defects, and (b) when either $\Theta\Sigma_y$ or $R_\pi$ symmetry is satisfied. $T_h=2500$ K and $T_c=1500$ K.}
	\label{Fig:Teff}
\end{figure}

The appearance of these surprising symmetry-protected currents in the nonthermal equilibrium situation can be explained in terms of the mode-dependent temperature $T^{\rm eff}_k$, Eq. \eqref{Teff}. Because the oscillatory behavior of $s_k$ and $r_k$ [the amplitude of a mode on the left ($s_k$) and right ($r_k$) edges strongly varies with the mode], $T^{\rm eff}_k$ presents a fluctuating profile resulting in an effective increment of noise and mixedness that generally removes edge currents. The system does not resist such a high degree of noise. However, under invariance by either $\Theta\Sigma_y$ or $R_{\pi}$, both coefficients are equal $s_k=r_k$, and therefore $T^{\rm eff}_k$ becomes independent of $s_k$ and $r_k$, Eq. \eqref{Teff}. This makes $T^{\rm eff}_k$ to be a monotonically increasing function of $\omega_k$, see Fig. \ref{Fig:Teff}, and there is a monotonic mode population with a similar pattern that in the thermal equilibrium situation: the more energy $\hbar\omega_k$, the less population in the mode, and edge currents are observed. Thus, the effect of disorder in the Liouvillian dynamics is minimal provided that the original symmetries are satisfied.

\section*{DISCUSSION}

The interaction of a bosonic topological system with one and two thermal baths presents a rich and new phenomenology. For one single bath, we find a wide range of temperatures where topological edge heat currents are present despite thermal fluctuations. For two baths, the edge heat current presents further remarkable properties: it is still robust with respect to defect perturbations as long as these defects respect certain discrete global symmetries. Moreover, one topological current flows against the natural arrow of the heat according to the second law (without implying any violation).

The \emph{selective} stability of the out-of-thermal-equilibrium current (SPNC) is due to a dissipative symmetry, or Liouvillian symmetry in the master equation \eqref{ME1}, not to any system Hamiltonian symmetry \eqref{H1}. Specifically, the system presents a new kind of protection, the \emph{dissipative symmetry-protection}, which minimizes the transition between the edge-conducting and the insulating phase in the steady state provided that some symmetries are satisfied. This is an open-system effect, similar to the usual Hamiltonian symmetry-protection of closed systems, where the transition probability between both states of matter is highly suppressed due to a symmetry property. This enforces the interest to study symmetry-protected topological ordered systems beyond the realm of Hamiltonian dynamics.

Moreover, note that in this case the property that an excitation or carrier has in order to overcome a defect is by no means understandable by some local argument. The excitation does or does not circulate around the defect depending on whether in other point of the lattice --which can be very far away-- there is another defect such that a global symmetry of the lattice is preserved. This manifests unquestionably the topological physics implied in this current.

Our results are based on a master equation formalism valid under the standard condition of weak coupling between system and reservoirs, and it is particularly suited to describe the steady state regime. For the sake of comparison, another approach based on a local formalism for the master equations of many body systems is given in the supplementary information document, but it does not lead to results (i)--(iv).

Although the study is carried out for the emblematic Hofstadter model of bosons, similar conclusions are drawn for the same class of topological insulators. We have not focused our attention on some specific platform, but master equations usually adequate very good to quantum optical systems. Therefore, set-ups based on optical lattices \cite{Bloch,Ketterle,Spielman} or photonic systems \cite{Hafezi1,Hafezi2,MIT2,Segev,MacDonald,MITReview} seem the most indicated to experimentally observe the effects here reported.

Finally, the edge current instability for a particular class of defect distributions is not only a novel, fundamental and intriguing effect, but may also impose some restrictions when manufacturing quantum Hall systems if we expect them to be stable when subject to some temperature gradient. This is of particular importance for example in the topological transport of photons \cite{Hafezi1,Hafezi2,MIT2,Segev,MacDonald,MITReview}, phonons \cite{PHE1,PHE2,PHE3,PHE4} or magnons \cite{MHE1,MHE2,MHE3} as their neutral charge prevents them to be displaced by applying electric fields.

\section*{METHODS}
\subsection*{Master Equation}
\label{app_A}

The bosonic Hofstadter Hamiltonian in general terms reads
\begin{equation}\label{SysHMM}
H_S=\sum_{x,y}\hbar\omega_0 a_{x,y}^\dagger a_{x,y}+V,
\end{equation}
with
\begin{align}\label{H1MM}
V=-\hbar J\sum_{x,y} a_{x+1,y}^\dagger a_{x,y} e^{i\theta_{x,y}^X} + a_{x,y+1}^\dagger a_{x,y} e^{i\theta_{x,y}^Y} +{\rm h.c.}
\end{align}
Here $a_{x,y}$ stands for the bosonic operator on the site $(x,y)$ of the $N\times N$ lattice and
\begin{align}
  \theta_{x,y}^X=\int_x^{x+1}\bm{A}\cdot d\bm{x}, \quad\text{and}\quad \theta_{x,y}^Y=\int_y^{y+1}\bm{A}\cdot d\bm{y},
\end{align}
where $\bm{A}$ denotes a gauge field.

After diagonalization, Eq. \eqref{SysHMM} yields $H_S=\sum_{k}\hbar \omega_k b_k^\dagger b_k$, where $\omega_k$ is the frequency of the normal mode $k$ and $a_{x,y}=\sum_k \psi_k(x,y) b_k$, with $\psi_k(x,y)$ the ``one-particle'' eigenfunctions. In terms of these eigenmodes the system-reservoir Hamiltonian can be written as
\begin{equation}\label{HSR3MM}
H_{SR}=\sum_{j,k} g_j \left(L_{j,k} b_{k}+L_{j,k}^\dagger b^\dagger_{k}\right) + g_j \left(R_{j,k} b_{k}+R_{j,k}^\dagger b^\dagger_{k}\right)
\end{equation}
with
\begin{align}
L_{j,k}:&=\sum_y \psi_k(1,y) (A_{j,y}+A^\dagger_{j,y}),\\
R_{j,k}:&=\sum_y \psi_k(N,y) (B_{j,y}+B^\dagger_{j,y}),
\end{align}
for operators of left and right reservoirs, respectively.

We obtain the Davies generator of the weak coupling limit \cite{Davies1,Davies2} by applying the standard procedure (technical details are in the supplementary information document). The subsequent master equation reads
\begin{widetext}
\begin{align}\label{ME1MM}
\frac{d\rho}{dt}=\mathcal{L}(\rho)=-\frac{i}{\hbar}[H_S,\rho] &+ \sum_k \gamma_k \big\{s_k  [\bar{n}_k(T_h)+1]+r_k [\bar{n}_k(T_c)+1]\big\} \Big(b_k \rho b_k^\dagger - \frac{1}{2}\{b_k^\dagger b_k,\rho\}\Big) \nonumber \\
&+ \sum_k \gamma_k \big[s_k \bar{n}_k(T_h)+r_k \bar{n}_k(T_c)\big] \Big(b_k^\dagger \rho b_k - \frac{1}{2}\{b_k b_k^\dagger,\rho\}\Big),
\end{align}
\end{widetext}
where $\bar{n}_k(T)=\{\exp[\hbar \omega_k/(k_B T)]-1\}^{-1}$ denotes the mean number of bosons with frequency $\omega_k$ and temperature $T$, and $\gamma_k$ is a constant that depends on the strength of the coupling via the spectral density $f(\omega)\sim\sum g_j^2\delta(\omega_j-\omega)$. For the sake of simplicity we shall assume the same decay rate for each eigenmode $\gamma_k=\gamma$, although this is not relevant to our conclusions. Furthermore the remaining constants in the equation are given by $s_k=\sum_{y=1}^N |\psi_k(1,y)|^2$ and $r_k=\sum_{y=1}^N |\psi_k(N,y)|^2$.

In the thermal equilibrium situation $T_h=T_c=T$, it is well-known (see e.g. \cite{Libros1,Libros2}) this master equation describes the dynamics of the system towards thermal equilibrium with the baths, so that the steady state obtained for long times is $\rho_\beta=\frac{e^{-\beta H_S}}{Z}$ with $\beta=1/(k_B T)$ and $Z=\Tr[\exp(-\beta H_S)]$.

For the general nonequilibrium case $T_h>T_c$, we note that the master equation \eqref{ME1MM} can be rewritten as
\begin{align}\label{ME3MM}
\frac{d\rho}{dt}=\mathcal{L}(\rho)=&-\frac{i}{\hbar}[H_S,\rho] \nonumber\\
&+ \sum_k \gamma [\bar{n}_k(T^{\rm eff}_k)+1] \Big(b_k \rho b_k^\dagger - \frac{1}{2}\{b_k^\dagger b_k,\rho\}\Big) \nonumber \\
&+ \sum_k \gamma \bar{n}_k(T^{\rm eff}_k)\Big(b_k^\dagger \rho b_k - \frac{1}{2}\{b_k b_k^\dagger,\rho\}\Big),
\end{align}
with an ``effective'' temperature $T^{\rm eff}_k$ depending on the mode and given by Eq. \eqref{Teff}. Therefore, we conclude that the steady state is of the form:
\begin{equation}\label{ssMM}
  \rho_{\rm ss}:=\lim_{t\rightarrow\infty}\rho(t)= W^{-1} \exp\left(-\sum_{k} \frac{\hbar \omega_k}{k_B T^{\rm eff}_k} b_k^\dagger b_k\right),
\end{equation}
with $W=\Tr\left[\exp\left(-\sum_{k} \frac{\hbar \omega_k}{k_B T^{\rm eff}_k} b_k^\dagger b_k\right)\right]$.

\subsection*{Current Operators}
\label{app_B}

External and internal currents are derived from continuity equations. For the external case, using the master equation \eqref{ME1MM} we have
\begin{align}\label{EqconMM}
\frac{d\langle H_S\rangle}{dt}&=\Tr\left(H_S\frac{d\rho}{dt}\right)=\Tr\left[H_S\mathcal{L}(\rho)\right]\nonumber\\
&=\langle \mathcal{L}^\sharp(H_S)\rangle=\langle \mathcal{J}_h\rangle+\langle \mathcal{J}_c\rangle,
\end{align}
where $\mathcal{J}_h$ ($\mathcal{J}_c$) is the current operator that describes the heat flux between system and hot (cold) bath, and $\mathcal{L}^\sharp$ denotes the Liouvillian at Eq.~\eqref{ME1MM} in the Heisenberg picture. On the other hand, since
\begin{equation}
\mathcal{L}^\sharp(H_S)=\mathcal{L}_c^\sharp(H_S)+\mathcal{L}_h^\sharp(H_S),
\end{equation}
with
\begin{align}
\mathcal{L}_{h}^\sharp(H_S)&:=\sum_{k} \gamma_k \big\{s_k  [\bar{n}_k(T_h)+1]\big\} \Big(b_k^\dagger H_S b_k - \frac{1}{2}\{b_k^\dagger b_k,H_S\}\Big) \nonumber \\
&+ \sum_k \gamma_k \big[s_k \bar{n}_k(T_h)\big] \Big(b_k H_S b_k^\dagger - \frac{1}{2}\{b_k b_k^\dagger,H_S\}\Big)\nonumber\\
&=-\hbar \sum_k \omega_k \gamma_k s_k [b_k^\dagger b_k- \bar{n}_k(T_h)],
\end{align}
and
\begin{align}
\mathcal{L}_{c}^\sharp(H_S)&:=\sum_{k} \gamma_k \big\{r_k  [\bar{n}_k(T_c)+1]\big\} \Big(b_k^\dagger H_S b_k - \frac{1}{2}\{b_k^\dagger b_k,H_S\}\Big) \nonumber \\
&+ \sum_k \gamma_k \big[r_k \bar{n}_k(T_c)\big] \Big(b_k H_S b_k^\dagger - \frac{1}{2}\{b_k b_k^\dagger,H_S\}\Big)\nonumber\\
 &=-\hbar \sum_k \omega_k \gamma_k r_k [b_k^\dagger b_k- \bar{n}_k(T_c)],
\end{align}
we identify the external currents as
\begin{align}
  \mathcal{J}_h:=-\hbar \sum_k \omega_k \gamma_k s_k [b_k^\dagger b_k- \bar{n}_k(T_h)],\\
  \mathcal{J}_c:=-\hbar \sum_k \omega_k \gamma_k r_k [b_k^\dagger b_k- \bar{n}_k(T_c)].
\end{align}
This identification is standard in the theory of open quantum systems, and it can be proven \cite{Spohn1} that the time-evolution described by the master equation \eqref{ME1MM} fulfills the entropy production inequality:
\begin{equation}\label{SprodMM}
\frac{d\mathcal{S}}{dt}-\frac{\langle\mathcal{J}_h\rangle}{T_h}-\frac{\langle\mathcal{J}_c\rangle}{T_c}\geq0,
\end{equation}
where $\mathcal{S}=-k_B\Tr(\rho\log\rho)$ is the thermodynamical entropy.

In order to derive internal currents we make use of the exact form of the continuity equation and the Davies' theorem \cite{Davies1}. Specifically, the exact equation for the population at the site $(x,y)$ is given by
\begin{align}\label{ExactContMM}
  \frac{d\langle a_{x,y}^\dagger a_{x,y}\rangle}{dt}=&\frac{i}{\hbar}\langle [H_S,a_{x,y}^\dagger a_{x,y}]\rangle+\frac{i}{\hbar}\langle [H_{SB},a_{x,y}^\dagger a_{x,y}]\rangle\nonumber\\
  =&-iJ\langle a_{x+1,y}^\dagger a_{x,y}e^{i\theta_{x,y}^X}-a_{x,y}^\dagger a_{x+1,y} e^{-i\theta_{x,y}^X}\rangle\nonumber\\
  &-iJ\langle a_{x-1,y}^\dagger a_{x,y} e^{-i\theta_{x,y}^X}-a_{x,y}^\dagger a_{x-1,y}e^{i\theta_{x,y}^X}\rangle\nonumber\\
  &-iJ\langle a_{x,y+1}^\dagger a_{x,y}e^{i\theta_{x,y}^Y}-a_{x,y}^\dagger a_{x,y+1}e^{-i\theta_{x,y}^Y} \rangle\nonumber\\
  &-iJ\langle a_{x,y-1}^\dagger a_{x,y}e^{-i\theta_{x,y}^Y}-a_{x,y}^\dagger a_{x,y-1}e^{i\theta_{x,y}^Y} \rangle\nonumber \\
  &+\frac{i}{\hbar}\langle [H_{SB},a_{x,y}^\dagger a_{x,y}]\rangle.
\end{align}
Then, the (internal) current operators are identified as:
\begin{align}\label{IntCurrent1MM}
  \mathcal{J}_{(\rightarrow x),y}&=iJ\big(a_{x,y}^\dagger a_{x-1,y}e^{i\theta_{x,y}^X}-a_{x-1,y}^\dagger a_{x,y} e^{-i\theta_{x,y}^X}\big),\\
  \mathcal{J}_{x,(\rightarrow y)}&=iJ\big(a_{x,y}^\dagger a_{x,y-1}e^{i\theta_{x,y}^Y}-a_{x,y-1}^\dagger a_{x,y}e^{-i\theta_{x,y}^Y}\big),\label{IntCurrent2MM}
\end{align}
where the subindex $(\rightarrow x)$ is a short notation for $(x-1 \rightarrow x)$. So that $\mathcal{J}_{(\rightarrow x),y}$ denotes the operator for the current leaving the site $(x-1,y)$ and entering in $(x,y)$. Similarly for $(\rightarrow y)$.

The term $\frac{i}{\hbar}\langle [H_{SB},a_{x,y}^\dagger a_{x,y}]\rangle$  in \eqref{ExactContMM} is not zero only for $x=1$ and $x=N$, and defines the exact external currents. Of course we cannot compute the exact time derivative $d\langle a_{x,y}^\dagger a_{x,y}\rangle/dt$; our approximation to it is given by the master equation \eqref{ME1MM}. In fact, the Davies theorem \cite{Davies1} asserts that the dissipative part of \eqref{ME1MM} is actually a weak coupling approximation of the term $\Tr_B\left(-i[H_{SB},\rho]\right)$ (see the supplementary information document for further details). This means that, in a weak coupling regime, it is consistent to take the above exact internal currents operators as internal current operators also in the master equation approximation.

In a Landau-type gauge taken throughout the main document, we write $\bm{A}=(-|\bm{B}|y,0,0)$ and internal currents take the form of Eqs. \eqref{IntCurrent1}--\eqref{IntCurrent2}, where $\alpha$ denotes the flux of the $\bm{B}$ field per unit cell.

More details about the derivation of the master equation and currents operators are provided in the supplementary information document.

\section*{Acknowledgments}

We acknowledge the Spanish MINECO grants FIS2015-67411, FIS2012-33152, and a ``Juan de la
Cierva-Incorporaci\'on'' research contract, the CAM research consortium QUITEMAD+ S2013/ICE-2801, and U.S. Army Research Office through grant W911NF-14-1-0103 for partial financial support.

\onecolumngrid

\newpage

\section*{SUPPLEMENTARY INFORMATION DOCUMENT}

\appendix

\setcounter{figure}{0}
\setcounter{equation}{0}
\renewcommand*{\thefigure}{S\arabic{figure}}
\renewcommand*{\theequation}{S\arabic{equation}}

\subsection*{I. Master Equation}
\label{app_A}

The bosonic Hofstadter Hamiltonian in general terms reads
\begin{equation}\label{SysHSM}
H_S=\sum_{x,y}\hbar\omega_0 a_{x,y}^\dagger a_{x,y}+V,
\end{equation}
with
\begin{align}\label{H1SM}
V=-\hbar J\sum_{x,y} a_{x+1,y}^\dagger a_{x,y} e^{i\theta_{x,y}^X} + a_{x,y+1}^\dagger a_{x,y} e^{i\theta_{x,y}^Y} +{\rm h.c.}
\end{align}
Here $a_{x,y}$ stands for the bosonic operator on the site $(x,y)$ of the $N \times N$ lattice and
\begin{align}
  \theta_{x,y}^X=\int_x^{x+1}\bm{A}\cdot d\bm{x}, \quad\text{and}\quad \theta_{x,y}^Y=\int_y^{y+1}\bm{A}\cdot d\bm{y},
\end{align}
where $\bm{A}$ denotes a gauge field.

The Hamiltonian Eq.~\eqref{SysHSM} can be written in matrix form as
\begin{equation}
H_S=\Psi^\dagger \mathbf{H_S} \Psi
\end{equation}
where $\Psi=(a_{1,1},\ldots,a_{N,1},a_{1,2},\ldots,a_{N,2},\ldots,a_{1,N},\ldots,a_{N,N})^{\rm t}$ and
\begin{equation}
\mathbf{H_S}=\hbar \begin{pmatrix}
M_1 						& I^\dagger_1					& 			&   \\
I_1						& M_2				& \ddots		&   \\
 						& \ddots				& \ddots		& I^\dagger_{N-1} \\
 						& 					& I_{N-1}			& M_N
\end{pmatrix},
\end{equation}
with $N\times N$ matrices
\begin{equation}
I_n=-J\begin{pmatrix}
e^{i \theta_{1,n}^Y}	&  	                &  \\
                 		& \ddots			&  \\
 						& 					&  e^{i \theta_{N,n}^Y}	
\end{pmatrix}
\quad\text{
and}\quad
M_n=\begin{pmatrix}
\omega_0					& -Je^{-i \theta_{1,n}^X}	& 			& \\
-Je^{i \theta_{1,n}^X}		& \ddots				& \ddots		& \\
 						& \ddots				& \ddots		& -Je^{-i \theta_{N-1,n}^X} \\
 						& 					& -Je^{i \theta_{N-1,n}^X}			& \omega_0	
\end{pmatrix},
\end{equation}
so that $\mathbf{H_S}$ is a $N^2\times N^2$ matrix.

\begin{figure}[t!]
	\includegraphics[width=0.5\columnwidth]{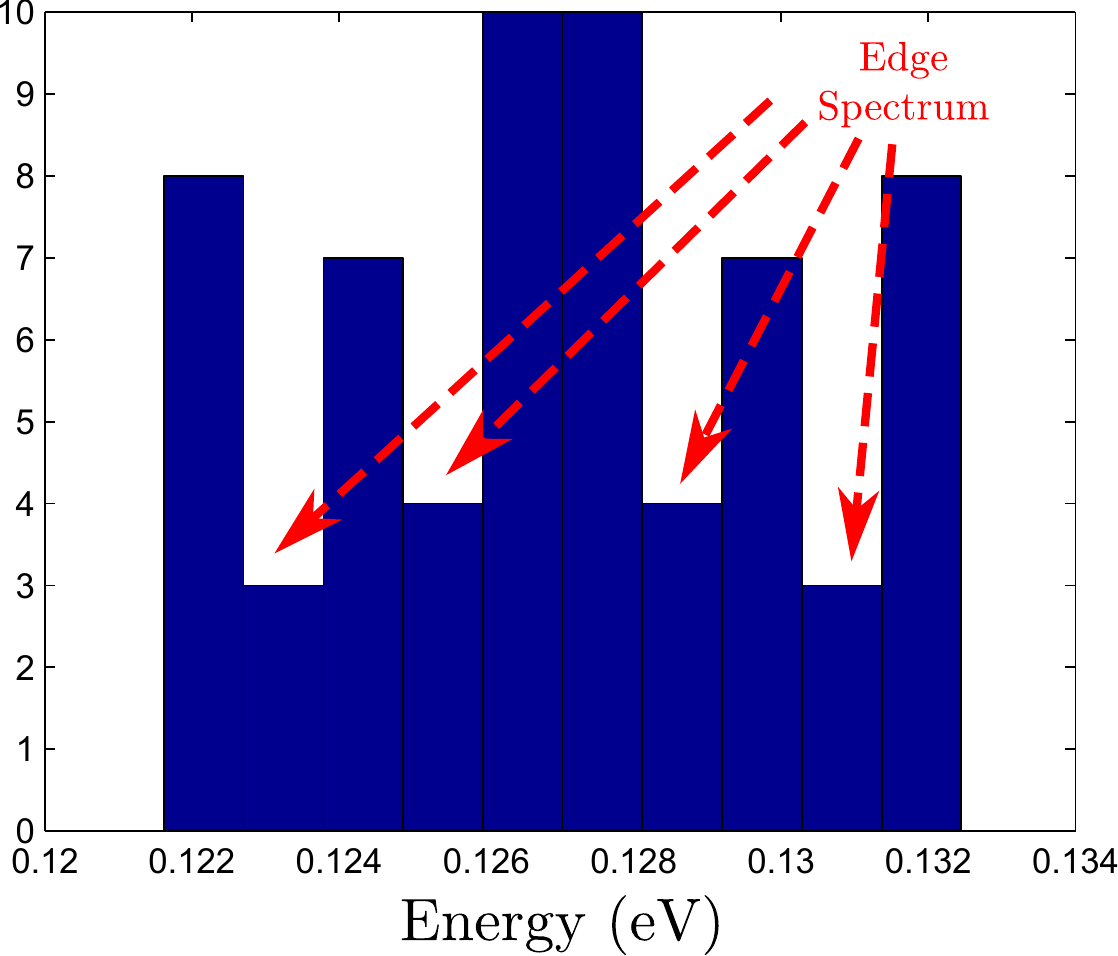}
	\caption{Density of states for an $8\times8$ lattice ($\omega_0=193\ {\rm THz}$ and $J=2.6\ {\rm THz}$). The four low density regions correspond to energies of edge eigenmodes.}
	\label{Fig:DOS}
\end{figure}

By diagonalizing the matrix $\mathbf{H_S}=\mathbf{U D U}^\dagger$ we obtain the spectrum of $H_S$ which can be visualized as the celebrated Hofstadter butterfly \cite{ButterflySM}. The eigenmodes of this Hamiltonian are given by $\bm{b}=\mathbf{U}^\dagger \Psi$, with $\bm{b}=(b_{1},b_{2},\ldots,b_{N^2})^{\rm t}$. In Fig.~\ref{Fig:DOS} we have represented the density of states for an $8\times8$ lattice, the low density regions correspond to energies associated with edge eigenmodes.

In terms of these eigenmodes the system-reservoir Hamiltonian yields
\begin{align}\label{HSR2SM}
H_{SR}&=\sum_{j,y} g_j (A_{j,y}+A^\dagger_{j,y})(a_{1,y}+a^\dagger_{1,y}) + g_j (B_{j,y}+B^\dagger_{j,y})(a_{N,y}+a^\dagger_{N,y})\nonumber \\
&=\sum_{j,y} g_j (A_{j,y}+A^\dagger_{j,y})\left(\sum_k u_{N(y-1)+1,k} b_{k}+u^\ast_{N(y-1)+1,k} b^\dagger_{k}\right) + g_j (B_{j,y}+B^\dagger_{j,y})\left(\sum_k u_{Ny,k} b_{k}+u^\ast_{Ny,k} b^\dagger_{k}\right),
\end{align}
where we have taken into account that the mode $a_{x,y}$ is allocated as the component $N(y-1)+x$ of the vector $\Psi$ and $u_{i,j}$ denotes the components of the unitary matrix $\mathbf{U}$. After some rearrangement, we write
\begin{equation}\label{HSR3SM}
H_{SR}=\sum_{j,k} g_j \left(L_{j,k} b_{k}+L_{j,k}^\dagger b^\dagger_{k}\right) + g_j \left(R_{j,k} b_{k}+R_{j,k}^\dagger b^\dagger_{k}\right)
\end{equation}
with
\begin{align}
L_{j,k}:=\sum_y u_{N(y-1)+1,k} (A_{j,y}+A^\dagger_{j,y}),\quad\text{and}\quad
R_{j,k}:=\sum_y u_{Ny,k} (B_{j,y}+B^\dagger_{j,y}),
\end{align}
for operators of left and right reservoirs, respectively.

Now, we obtain the Davies generator of the weak coupling limit \cite{Davies} by applying the standard procedure (see, for instance, \cite{Libros}). Note that there are not mixed terms between different site reservoirs in the correlation functions, so the second order dissipator becomes
\begin{align}
\mathcal{D}(\rho)= \sum_k  \int_{-\infty}^\infty d\tau  \sum_{j} &g_j^2 e^{i\omega_k\tau} \left(\langle\tilde{L}_{j,k}^\dagger(\tau) L_{j,k}\rangle_{\beta_h}+\langle\tilde{R}_{j,k}^\dagger(\tau) R_{j,k}\rangle_{\beta_c}\right) [b_k \rho b_k^\dagger - \frac{1}{2}\{b_k^\dagger b_k,\rho\}]\nonumber \\
+ &g_j^2 e^{-i\omega_k\tau} \left(\langle\tilde{L}_{j,k}(\tau) L_{j,k}^\dagger\rangle_{\beta_h}+\langle\tilde{R}_{j,k}(\tau) R_{j,k}^\dagger\rangle_{\beta_c}\right) [b_k^\dagger \rho b_k - \frac{1}{2}\{b_k b_k^\dagger,\rho\}],
\end{align}
where $\tilde{L}_{j,k}(\tau)=\sum_y u_{N(y-1)+1,k} [A_{j,y}\exp(-i\nu_{j,y} \tau)+A^\dagger_{j,y}\exp(i\nu_{j,y} \tau)]$, and (similarly for) $\tilde{R}_{j,k}(\tau)$, are the operators $L_{j,k}$ and $R_{j,k}$ in the interaction picture ($\nu_{j,y}$ is the frequency of the bath mode $A_{j,y}$), $\omega_k$ is the frequency associated to the eigenmode $b_k$, and $\beta_{h,c}=1/(k_B T_{h,c})$ are inverse temperatures of hot and cold baths. Following the usual steps for the derivation \cite{Libros}, we obtain the following master equation:
\begin{align}\label{ME1SM}
\frac{d\rho}{dt}=\mathcal{L}(\rho)=-\frac{i}{\hbar}[H_S,\rho] &+ \sum_k \gamma_k \big\{s_k  [\bar{n}_k(T_h)+1]+r_k [\bar{n}_k(T_c)+1]\big\} \Big(b_k \rho b_k^\dagger - \frac{1}{2}\{b_k^\dagger b_k,\rho\}\Big) \nonumber \\
&+ \sum_k \gamma_k \big[s_k \bar{n}_k(T_h)+r_k \bar{n}_k(T_c)\big] \Big(b_k^\dagger \rho b_k - \frac{1}{2}\{b_k b_k^\dagger,\rho\}\Big),
\end{align}
where $\bar{n}_k(T)=\{\exp[\hbar \omega_k/(k_B T)]-1\}^{-1}$ denotes the mean number of bosons with frequency $\omega_k$ and temperature $T$, and $\gamma_k$ is a constant that depends on the strength of the coupling via the spectral density $f(\omega)\sim\sum g_j^2\delta(\omega_j-\omega)$. For the sake of simplicity we have assumed in the main text the same decay rate for each eigenmode $\gamma_k=\gamma$, although this is not relevant to our conclusions. Furthermore the constants $s_k$ and $r_k$ are related to the matrix $\mathbf{U}$ via:
\begin{align}
s_k&=\sum_{y=1}^N [\mathbf{U}^\dagger]_{k,N(y-1)+1}[\mathbf{U}]_{N(y-1)+1,k}=\sum_{y=1}^N |u_{N(y-1)+1,k}|^2,\label{sdef1SM}\\
r_k&=\sum_{y=1}^N [\mathbf{U}^\dagger]_{k,Ny}[\mathbf{U}]_{Ny,k}=\sum_{x=1}^N |u_{Ny,k}|^2.\label{rdef1SM}
\end{align}
Since the columns of the matrix $\textbf{U}$ are the coordinates in real space of ``one-particle" wavefunctions $\psi_k(x,y)$, after reordering, we obtain
\begin{align}\label{srdef2SM}
s_k=\sum_{y=1}^N |\psi_k(1,y)|^2,\quad \text{and} \quad r_k=\sum_{y=1}^N |\psi_k(N,y)|^2.
\end{align}

In the thermal equilibrium situation $T_h=T_c=T$, the master equation \eqref{ME1SM} becomes
\begin{align}\label{ME2SM}
\frac{d\rho}{dt}=\mathcal{L}(\rho)=-\frac{i}{\hbar}[H_S,\rho] &+ \sum_k \bar{\gamma}_k [\bar{n}_k(T)+1] \Big(b_k \rho b_k^\dagger - \frac{1}{2}\{b_k^\dagger b_k,\rho\}\Big) \nonumber \\
&+ \sum_k \bar{\gamma}_k \bar{n}_k(T)\Big(b_k^\dagger \rho b_k - \frac{1}{2}\{b_k b_k^\dagger,\rho\}\Big),
\end{align}
with $\bar{\gamma}_k=\gamma (s_k+r_k)$. This equation describes the dynamics of the system towards thermal equilibrium with the baths, so that the steady state obtained for long times is the Gibbs state at the same temperature as the baths:
\begin{equation}
\rho_{\rm ss}:=\lim_{t\rightarrow\infty}\rho(t)=\rho_\beta=\frac{e^{-\beta H_S}}{Z},
\end{equation}
with $\beta=1/(k_B T)$ and $Z=\Tr[\exp(-\beta H_S)]$.

In the general nonequilibrium case $T_h>T_c$, it is usually involved to obtain the steady state at the stationary limit. However, in this case we can find it by noting that the master equation \eqref{ME1SM} can be rewritten as
\begin{align}\label{ME3SM}
\frac{d\rho}{dt}=\mathcal{L}(\rho)=-\frac{i}{\hbar}[H_S,\rho] &+ \sum_k \gamma [\bar{n}_k(T^{\rm eff}_k)+1] \Big(b_k \rho b_k^\dagger - \frac{1}{2}\{b_k^\dagger b_k,\rho\}\Big) \nonumber \\
&+ \sum_k \gamma \bar{n}_k(T^{\rm eff}_k)\Big(b_k^\dagger \rho b_k - \frac{1}{2}\{b_k b_k^\dagger,\rho\}\Big),
\end{align}
with some ``effective'' temperature $T^{\rm eff}_k$ depending on the mode:
\begin{equation}\label{TeffSM}
  T^{\rm eff}_k:=\frac{\hbar \omega_k}{k_B\log\left\{\frac{\exp\big(\tfrac{\hbar \omega_k}{k_B T_h}\big)\big[\exp\big(\tfrac{\hbar \omega_k}{k_B T_c}\big)-1\big] s_k + \exp\big(\tfrac{\hbar \omega_k}{k_B T_c}\big)\big[\exp\big(\tfrac{\hbar \omega_k}{k_B T_h}\big)-1\big] r_k}{\big[\exp\big(\tfrac{\hbar \omega_k}{k_B T_c}\big)-1\big] s_k + \big[\exp\big(\tfrac{\hbar \omega_k}{k_B T_h}\big)-1\big] r_k}\right\}}.
\end{equation}
Therefore, the master equation \eqref{ME3SM} is the same as the one describing the dynamics of a collection of $N^2$ independent modes with different frequencies interacting with $N^2$ thermal baths with different temperatures $T^{\rm eff}_k$. Hence, we conclude that the steady state is of the form:
\begin{equation}\label{ssSM}
  \rho_{\rm ss}:=\lim_{t\rightarrow\infty}\rho(t)= W^{-1} \exp\left(-\sum_{k} \frac{\hbar \omega_k}{k_B T^{\rm eff}_k} b_k^\dagger b_k\right),
\end{equation}
with $W=\Tr\left[\exp\left(-\sum_{k} \frac{\hbar \omega_k}{k_B T^{\rm eff}_k} b_k^\dagger b_k\right)\right]$.

\subsection*{II. Current Operators}
\label{app_B}

External and internal currents are derived from continuity equations. For the external case, using the master equation \eqref{ME1SM} we have
\begin{equation}\label{EqconSM}
\frac{d\langle H_S\rangle}{dt}=\Tr\left(H_S\frac{d\rho}{dt}\right)=\Tr\left[H_S\mathcal{L}(\rho)\right]=\langle \mathcal{L}^\sharp(H_S)\rangle=\langle \mathcal{J}_h\rangle+\langle \mathcal{J}_c\rangle,
\end{equation}
where $\mathcal{J}_h$ ($\mathcal{J}_c$) is the current operator that describes the heat flux between system and hot (cold) bath, and $\mathcal{L}^\sharp$ denotes the Liouvillian at Eq.~\eqref{ME1SM} in the Heisenberg picture. Note that in the usual convention a positive current (meaning an increment of $\langle H_S\rangle$) is associated to energy flowing from outside to the system, whereas a negative current describes energy flowing from system to outside. On the other hand, since
\begin{equation}
\mathcal{L}^\sharp(H_S)=\mathcal{L}_c^\sharp(H_S)+\mathcal{L}_h^\sharp(H_S),
\end{equation}
with
\begin{align}
\mathcal{L}_{h}^\sharp(H_S)&:=\sum_{k} \gamma_k \big\{s_k  [\bar{n}_k(T_h)+1]\big\} \Big(b_k^\dagger H_S b_k - \frac{1}{2}\{b_k^\dagger b_k,H_S\}\Big) \nonumber \\
&+ \sum_k \gamma_k \big[s_k \bar{n}_k(T_h)\big] \Big(b_k H_S b_k^\dagger - \frac{1}{2}\{b_k b_k^\dagger,H_S\}\Big)=-\hbar \sum_k \omega_k \gamma_k s_k [b_k^\dagger b_k- \bar{n}_k(T_h)],
\end{align}
and
\begin{align}
\mathcal{L}_{c}^\sharp(H_S)&:=\sum_{k} \gamma_k \big\{r_k  [\bar{n}_k(T_c)+1]\big\} \Big(b_k^\dagger H_S b_k - \frac{1}{2}\{b_k^\dagger b_k,H_S\}\Big) \nonumber \\
&+ \sum_k \gamma_k \big[r_k \bar{n}_k(T_c)\big] \Big(b_k H_S b_k^\dagger - \frac{1}{2}\{b_k b_k^\dagger,H_S\}\Big)=-\hbar \sum_k \omega_k \gamma_k r_k [b_k^\dagger b_k- \bar{n}_k(T_c)],
\end{align}
we identify the currents as
\begin{align}
  \mathcal{J}_h:=-\hbar \sum_k \omega_k \gamma_k s_k [b_k^\dagger b_k- \bar{n}_k(T_h)],\\
  \mathcal{J}_c:=-\hbar \sum_k \omega_k \gamma_k r_k [b_k^\dagger b_k- \bar{n}_k(T_c)].
\end{align}
This identification is standard in the theory of open quantum systems, and it can be proven \cite{Spohn1} that the time-evolution described by the master equation \eqref{ME1SM} fulfills the entropy production inequality:
\begin{equation}\label{SprodSM}
\frac{d\mathcal{S}}{dt}-\frac{\langle\mathcal{J}_h\rangle}{T_h}-\frac{\langle\mathcal{J}_c\rangle}{T_c}\geq0,
\end{equation}
where $\mathcal{S}=-k_B\Tr(\rho\log\rho)$ is the thermodynamical entropy. In the stationary regime, $t\rightarrow\infty$, the system approaches some steady state $\lim_{t\rightarrow\infty}\rho(t)=\rho_{\rm ss}$, so that $\frac{d\langle H_S\rangle}{dt}=0$ and $\tfrac{d\mathcal{S}_{\rm ss}}{dt}=0$, and Eqs. \eqref{EqconSM} and \eqref{SprodSM} yield
\begin{align}
&\langle\mathcal{J}_h\rangle_{\rm ss}+\langle\mathcal{J}_c\rangle_{\rm ss}=0,\\
&\frac{\langle\mathcal{J}_h\rangle_{\rm ss}}{T_h}+\frac{\langle\mathcal{J}_c\rangle_{\rm ss}}{T_c}\leq0.
\end{align}
Since $T_h>T_c$, the above relations impose that $\langle\mathcal{J}_h\rangle_{\rm ss}=-\langle\mathcal{J}_c\rangle_{\rm ss}>0$. This is in agreement with the second law of thermodynamics, in particular in the Clausius formulation as if no work is performed on system, the current goes from hot to cold bath \cite{Clausius}. Note that if all baths are at the same temperature, $T_h=T_c=T$, once the stationary limit has been reached, the net external current between system and baths becomes obliviously zero.

Coming back to the system of our interest, we can split the current operator corresponding to hot $\mathcal{J}_h$ and cold $\mathcal{J}_c$ baths as a sum of currents operators for individual baths. Specifically, using Eq.~\eqref{srdef2SM},
\begin{equation}
  \mathcal{J}_h=-\hbar \sum_k \omega_k \gamma_k s_k [b_k^\dagger b_k- \bar{n}_k(T_h)]=-\hbar \sum_{y=1}^N \sum_k \omega_k \gamma_k  |\psi_k(1,y)|^2 [b_k^\dagger b_k- \bar{n}_k(T_h)]=\sum_{y=1}^N\mathcal{J}_h^y,
\end{equation}
where $\mathcal{J}_h^y=-\hbar \sum_k \omega_k \gamma_k |\psi_k(1,y)|^2 [b_k^\dagger b_k- \bar{n}_k(T_h)]$ is the current operator accounting for the energy flow between the hot bath at position $y$ and the system. Similarly, $\mathcal{J}_c=\sum_{y=1}^N\mathcal{J}_c^y$, with $\mathcal{J}_c^y=-\hbar \sum_k \omega_k \gamma_k |\psi_k(N,y)|^2 [b_k^\dagger b_k- \bar{n}_k(T_c)]$.

To obtain $\langle \mathcal{J}_h^y\rangle_{\rm ss}$ and $\langle \mathcal{J}_c^y\rangle_{\rm ss}$, we solve the dynamical equation for $\langle b_k^\dagger b_k\rangle$:
\begin{equation}
  \frac{d\langle b_k^\dagger b_k\rangle}{dt}=-\gamma  [(s_k+r_k)\langle b_k^\dagger b_k\rangle- s_k\bar{n}_k(T_h)- r_k\bar{n}_k(T_c)],
\end{equation}
obtaining
\begin{equation}
\langle b_k^\dagger b_k\rangle(t)=e^{-\gamma(s_k+r_k)t}\langle b_k^\dagger b_k\rangle(0)+\frac{[1-e^{-\gamma(s_k+r_k)t}]}{s_k + r_k} [s_k\bar{n}_k(T_h)+ r_k\bar{n}_k(T_c)],
\end{equation}
so that
\begin{equation}
\langle b_k^\dagger b_k\rangle_{\rm ss}=\lim_{t\rightarrow\infty}\langle b_k^\dagger b_k\rangle(t)=\frac{s_k\bar{n}_k(T_h)+ r_k\bar{n}_k(T_c)}{s_k + r_k}.
\end{equation}
Therefore, the external currents at the stationary state are
\begin{align}
  \langle\mathcal{J}_h^y\rangle_{\rm ss}&=\hbar \sum_k \omega_k \gamma_k r_k |\psi_k(1,y)|^2 \left[\frac{\bar{n}_k(T_h)-\bar{n}_k(T_c)}{s_k + r_k}\right],\\
  \langle\mathcal{J}_c^y\rangle_{\rm ss}&=\hbar \sum_k \omega_k \gamma_k s_k |\psi_k(N,y)|^2  \left[\frac{\bar{n}_k(T_c)-\bar{n}_k(T_h)}{s_k + r_k}\right].
\end{align}

In order to derive internal currents we make use of the exact form of the continuity equation and the Davies' theorem \cite{Davies}. Specifically, the exact equation for the population at the site $(x,y)$ is given by
\begin{align}\label{ExactContSM}
  \frac{d\langle a_{x,y}^\dagger a_{x,y}\rangle}{dt}=&\frac{i}{\hbar}\langle [H_S,a_{x,y}^\dagger a_{x,y}]\rangle+\frac{i}{\hbar}\langle [H_{SB},a_{x,y}^\dagger a_{x,y}]\rangle\nonumber\\
  =&-iJ\langle a_{x+1,y}^\dagger a_{x,y}e^{i\theta_{x,y}^X}-a_{x,y}^\dagger a_{x+1,y} e^{-i\theta_{x,y}^X}\rangle-iJ\langle a_{x-1,y}^\dagger a_{x,y} e^{-i\theta_{x,y}^X}-a_{x,y}^\dagger a_{x-1,y}e^{i\theta_{x,y}^X}\rangle\nonumber\\
  &-iJ\langle a_{x,y+1}^\dagger a_{x,y}e^{i\theta_{x,y}^Y}-a_{x,y}^\dagger a_{x,y+1}e^{-i\theta_{x,y}^Y} \rangle-iJ\langle a_{x,y-1}^\dagger a_{x,y}e^{-i\theta_{x,y}^Y}-a_{x,y}^\dagger a_{x,y-1}e^{i\theta_{x,y}^Y} \rangle\nonumber \\
  &+\frac{i}{\hbar}\langle [H_{SB},a_{x,y}^\dagger a_{x,y}]\rangle.
\end{align}
Then, the (internal) current operators are identified as:
\begin{align}\label{IntCurrent1SM}
  \mathcal{J}_{(\rightarrow x),y}&=iJ\big(a_{x,y}^\dagger a_{x-1,y}e^{i\theta_{x,y}^X}-a_{x-1,y}^\dagger a_{x,y} e^{-i\theta_{x,y}^X}\big),\\
  \mathcal{J}_{x,(\rightarrow y)}&=iJ\big(a_{x,y}^\dagger a_{x,y-1}e^{i\theta_{x,y}^Y}-a_{x,y-1}^\dagger a_{x,y}e^{-i\theta_{x,y}^Y}\big),\label{IntCurrent2SM}
\end{align}
where the subindex $(\rightarrow x)$ is a short notation for $(x-1 \rightarrow x)$. So that $\mathcal{J}_{(\rightarrow x),y}$ denotes the operator for the current leaving the site $(x-1,y)$ and entering in $(x,y)$. Similarly for $(\rightarrow y)$.

The term $\frac{i}{\hbar}\langle [H_{SB},a_{x,y}^\dagger a_{x,y}]\rangle$  in \eqref{ExactContSM} is not zero only for $x=1$ and $x=N$, and defines the exact external currents. Of course we cannot compute the exact time derivative $d\langle a_{x,y}^\dagger a_{x,y}\rangle/dt$; our approximation to it is given by the master equation \eqref{ME1SM}. However, if one tries to define internal currents directly from Eq.~\eqref{ME1SM}, one finds a problem because the approximation introduces fictitious dissipative couplings among all oscillators. Nevertheless, the Davies theorem \cite{Davies} asserts that the dissipative part of \eqref{ME1SM} is actually a weak-coupling approximation of the term $\Tr_B\left(-i[H_{SB},\rho]\right)$ \cite{footnote1}. This suggests that, in a weak coupling regime, it is consistent to take the above exact internal currents operators as internal current operators also in the master equation approximation. In this manner, the terms $\frac{i}{\hbar}\langle [H_{SB},a_{1,y}^\dagger a_{1,y}]\rangle$ are intended to be described by $\mathcal{J}_h^y$ as defined above.

In a Landau-type gauge taken throughout the main document, we write $\bm{A}=(-|\bm{B}|y,0,0)$ and internal currents take the form
\begin{align}
\mathcal{J}_{(\rightarrow x),y}&:=iJ\big(a_{x,y}^\dagger a_{x-1,y}e^{-2\pi \alpha i y}-a_{x-1,y}^\dagger a_{x,y} e^{2\pi \alpha i y}\big),\\
  \mathcal{J}_{x,(\rightarrow y)}&:=iJ\big(a_{x,y}^\dagger a_{x,y-1}-a_{x,y-1}^\dagger a_{x,y}\big),
\end{align}
where $\alpha$ stands for the flux of $\bm{B}$ per plaquette. These equations, in terms of normal modes, become
\begin{align}
  \mathcal{J}_{(\rightarrow x),y}&=iJ\sum_{k,k'}u^\ast_{N(y-1)+x,k}u_{N(y-1)+(x-1),k'}e^{-2\pi\alpha i y}b_k^\dagger b_{k'} + {\rm h.c.},\\
  \mathcal{J}_{x,(\rightarrow y)}&=iJ\sum_{k,k'}u^\ast_{N(y-1)+x,k}u_{N(y-2)+x,k'}b_k^\dagger b_{k'} + {\rm h.c.}
\end{align}
At the stationary limit the steady state is given by Eq.~\eqref{ssSM} and we obtain
\begin{align}
  \langle \mathcal{J}_{(\rightarrow x),y}\rangle &=\frac{J}{2} {\rm Im}\sum_{k}u^\ast_{N(y-1)+(x-1),k}u_{N(y-1)+x,k}e^{2\pi\alpha i y}\langle b_k^\dagger b_{k}\rangle \nonumber \\
  &=\frac{J}{2} \sum_{k} \bar{n}_k(T^{\rm eff}_k){\rm Im}[u^\ast_{N(y-1)+(x-1),k}u_{N(y-1)+x,k}e^{2\pi\alpha i y}],\\
  \langle \mathcal{J}_{x,(\rightarrow y)}\rangle &=\frac{J}{2} {\rm Im}\sum_{k}u^\ast_{N(y-2)+x,k}u_{N(y-1)+x,k}\langle b_k^\dagger b_{k}\rangle\nonumber \\
  &=\frac{J}{2} \sum_{k} \bar{n}_k(T^{\rm eff}_k) {\rm Im}[u^\ast_{N(y-2)+x,k}u_{N(y-1)+x,k}].
\end{align}

Note that these internal currents describe the flux of carriers or quanta per time, we do not aim to associate any specific energy with the current from some site to its adjacent independently of what normal mode is excited in the lattice.

\begin{figure}[t]
	\includegraphics[width=0.7\textwidth]{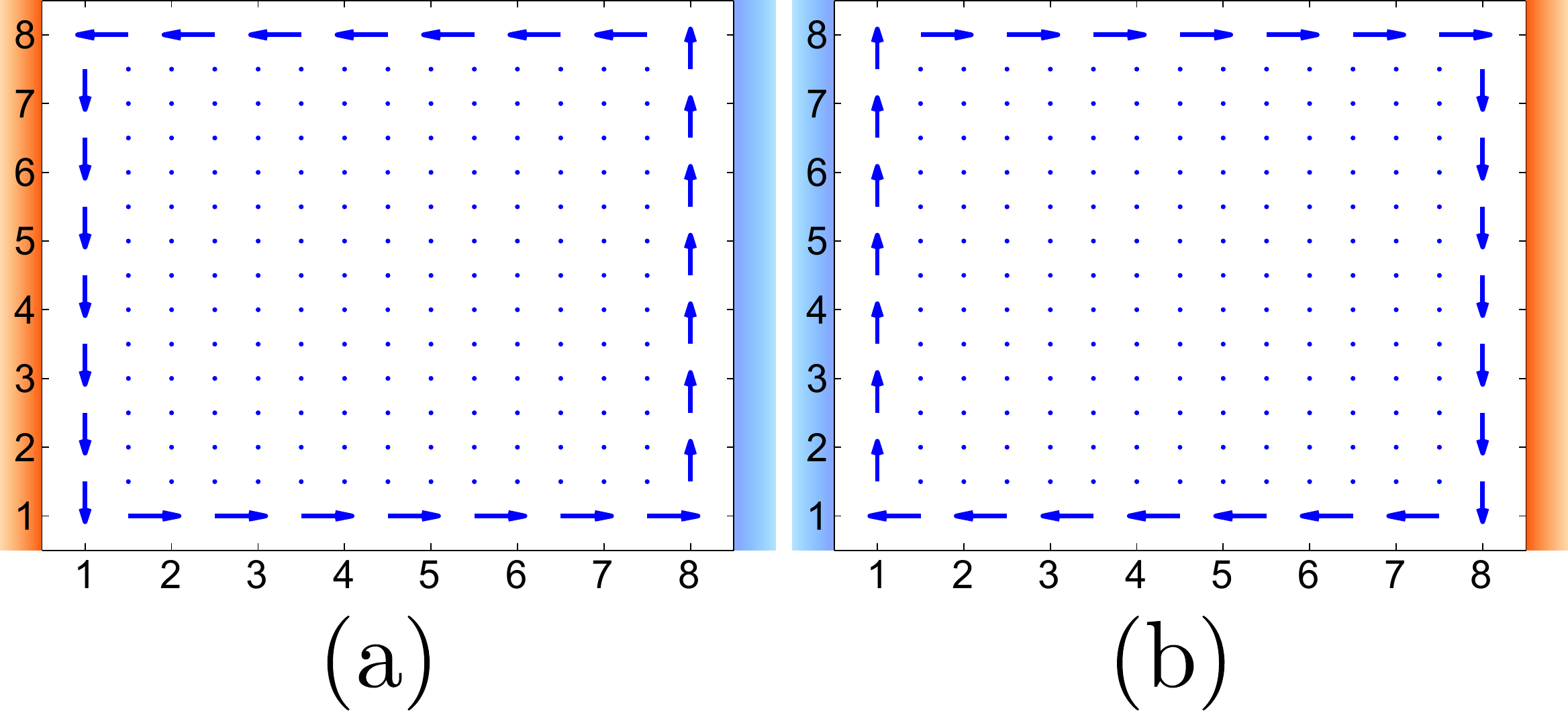}
	\caption{Current patterns for an out-of-thermal-equilibrium situation. In (a) the lattice is in contact with a bath at $T_h=2500\ {\rm K}$ on the left side and with a bath at $T_c=1500\ {\rm K}$ of the right side. Note the anomalous current on the top edge in opposite direction to the heat flow. The inverted situation along the direction of the temperature gradient is depicted in (b) showing a reversed current.}
	\label{Fig:NONThermal1SM}
\end{figure}

In addition, note also that if $s_k=r_k$, $T^{\rm eff}_k$ is invariant under the exchange $T_h\leftrightarrow T_c$, (because $s_k=r_k$). This, in particular, implies that the internal currents do not change, neither their absolute value nor their sign, under the exchange $T_h\leftrightarrow T_c$. This may be surprising at first sight, but it is due to the fact that a simple exchange of $T_h\leftrightarrow T_c$ can be seen as lattice rotation $R_{\pi}$, because in that case left and right temperature are exchanged and the magnetic field and lattice properties remain invariant. Under such a rotation, it seems natural that currents do not change their value as they are chiral. However, under a reflection along the temperature gradient direction $x\leftrightarrow -x$, the temperatures are exchanged $T_h\leftrightarrow T_c$ but also the magnetic field changes $\bm{B} \leftrightarrow -\bm{B}$. In this situation the currents indeed change their sign. This is showed in Fig.~\ref{Fig:NONThermal1SM}.

\begin{figure}[t]
	\includegraphics[width=0.6\textwidth]{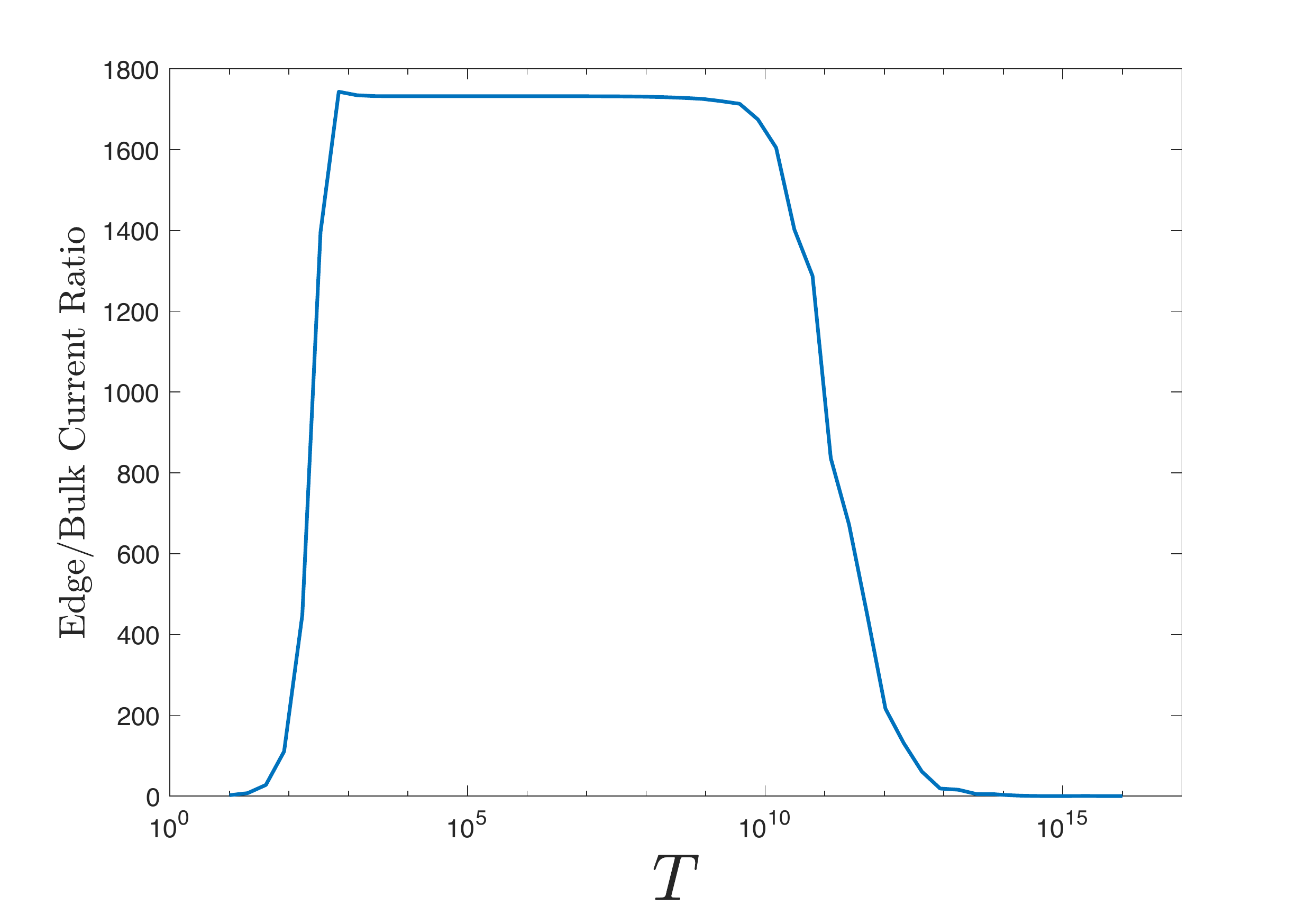}
	\caption{Ratio between edge and bulk current as a function of the temperature. We have taken a $50\times 50$ array with the same values for sites energies and couplings as in the main text.}
	\label{Fig:RatioSM}
\end{figure}

Finally, in Fig. \ref{Fig:RatioSM} we depict the edge/bulk current ratio as a function of the temperature for the numerical parameters taken throughout the manuscript. We can distinguish the three phases in Fig. 2 of the main text. For low and high temperatures the bulk current is similar to (or higher than) the edge current. However, for a large intermediate range of temperatures the system remains in a phase with high edge current concentration.

\subsection*{III. Master Equation in the `Local' Approach}
\label{app_C}

In the analysis of internal currents it is also common to consider the sometimes refereed to as the ``local'' approach to the master equation. In this approach the master equation for our system reads
\begin{align}\label{ME4SM}
\frac{d\rho}{dt}=\mathcal{L}(\rho)=-\frac{i}{\hbar}[H_S,\rho] &+ \gamma\sum_{y=1}^N  [\bar{n}(T_h)+1]\Big(a_{1,y} \rho a_{1,y}^\dagger - \frac{1}{2}\{a_{1,y}^\dagger a_{1,y},\rho\}\Big) \nonumber \\
&+  \bar{n}(T_h)\Big(a_{1,y}^\dagger \rho a_{1,y} - \frac{1}{2}\{a_{1,y} a_{1,y}^\dagger,\rho\}\Big) \nonumber\\
&+  [\bar{n}(T_c)+1]\Big(a_{N,y} \rho a_{N,y}^\dagger - \frac{1}{2}\{a_{N,y}^\dagger a_{N,y},\rho\}\Big) \nonumber \\
&+ \bar{n}(T_c)\Big(a_{N,y}^\dagger \rho a_{N,y} - \frac{1}{2}\{a_{N,y} a_{N,y}^\dagger,\rho\}\Big).
\end{align}
with $\bar{n}(T)=\{\exp[\hbar \omega_0/(k_B T)]-1\}^{-1}$.

This master equation corresponds to the introduction of dissipation in a kind of ``adiabatic'' way; by assuming that the dissipative process is not affected when changing $J$ from 0 to some small parameter (see discussion in \cite{Libros,NJPUlm}). So, this master equation is expected to provide a good description of the dynamics as long as $J$ is small in comparison with the typical time scale of the system when $J=0$. In the stationary regime at the limit $t\rightarrow\infty$, this equation is not expected to be a good description and it presents some problems from a thermodynamical point of view. Namely,
\begin{enumerate}
  \item For $T_h=T_c=T$, the state at the long time limit provided by this master equation is not a Gibbs state. Of course, if $J$ is small enough both states are very close \cite{NJPUlm}, however this does not guaranty a correct thermodynamic description \cite{OneTLocalApproach}.
  \item For $T_h>T_c$, external currents may violate the second law of thermodynamics \cite{LeviKosloff}, accounting for an unphysical heat flow from cold to hot bath.
\end{enumerate}
Yet, for the sake of comparison we have solved this equation \eqref{ME4SM}: no chiral currents are found, and the pattern is independent of the values of $\alpha$. This is in agreement with the results reported on \cite{Poletti} for a lattice with two rows (a ladder) where no chiral current was obtained.


\begin{thebibliography}{99}
\bibitem[S6]{footnote1} Davies' theorem states $\Tr_B[\exp(-iHt)\rho_S\otimes\rho_B\exp(iHt)]\approx\exp(\mathcal{L}t)\rho_S$. By differentiation $\Tr_B\{-i[H,\exp(-iHt)\rho_S\otimes\rho_B\exp(iHt)]\}\approx\mathcal{L}[\exp(\mathcal{L}t)\rho_S]$. Since $H=H_S+H_B+H_{SB}$, $\mathcal{L}(\cdot)=-i[H_S,\cdot]+\mathcal{D}(\cdot)$ and
    \begin{equation}
    \Tr_B\{-i[H_S,\exp(-iHt)\rho_S\otimes\rho_B\exp(iHt)]\}=-i[H_S,\Tr_B\{\exp(-iHt)\rho_S\otimes\rho_B\exp(iHt)\}]\approx-i[H_S,\exp(\mathcal{L}t)\rho_S],
    \end{equation}
    ones concludes $\Tr_B\{-i[H_{SB},\exp(-iHt)\rho_S\otimes\rho_B\exp(iHt)]\}\approx \mathcal{D}[\exp(\mathcal{L}t)\rho_S]$.
\end{thebibliography}
\end{document}